\documentstyle[aps,prl,twocolumn]{revtex}
\draft
\begin{document}   

\include{psfig}
\draft

\title{Extension of the Fluctuation-Dissipation theorem 
to the physical aging of a model glass-forming liquid}
 

\author{Francesco Sciortino and Piero Tartaglia}

\address{Dipartimento di Fisica, Universit\'a di Roma
"La Sapienza" and Istituto Nazionale per la Fisica della Materia,
Piazzale Aldo Moro 2, I-00185, Roma, Italy}

\date{Draft: \today}
\maketitle

\begin{abstract}
We present evidence in favor of the possibility of treating
an out-of-equilibrium supercooled 
simple liquid as a system in quasi-equilibrium.
Two different temperatures, one controlled by the external bath and one 
internally selected by the system characterize the quasi-equilibrium state.
The value of the
internal temperature is explicitly calculated within 
the inherent structure thermodynamic formalism.
We find that 
the internal temperature controls the  relation between the response
to an external perturbation and the long-time 
decay of fluctuations in the liquid.
\end{abstract}

In the last decade, several efforts have been devoted to the
understanding of the glass-transition
phenomenon\cite{debenedetti,review,goetze}, one of the open
fundamental problems of condensed matter.
In particular, the thermodynamics of supercooled liquids and the
relations between dynamics and thermodynamics have received
considerable interest.  Novel
approaches\cite{parisi,franz1,coniglio,teo,speedy} and detailed analyses of
computer generated configurations of supercooled liquids
\cite{heuer,thomas,barbara,prlentro,srinew}
are supporting and formalizing the picture --- rooted in old
ideas\cite{goldstein} --- of a glass as a system trapped in one local
free energy basin. Correspondingly, the equilibrium dynamics of deep
supercooled liquids is interpreted as motion among different
basins.

In this Letter, we show 
that the thermodynamic approach also allows us to interpret
the behavior of supercooled model liquids under non-equilibrium
(aging) situations.  The model liquid we study is the Lennard Jones
binary (80:20) mixture (BMLJ) which has been extensively investigated
in the past\cite{kob,system}.  The formalism we adopt is the so-called
inherent structure ($IS$) formalism\cite{stillinger}, which focus on
the local minima of the potential energy surface and on the
corresponding basins of attraction.  In this formalism, each point in
the 3$N$-dimensional configuration space is unambiguously associated
to the basin of attraction of the closest $IS$.  The configuration
space is partitioned in a sum of basins (usually labeled according to
the value $e_{IS}$ of the potential energy in the $IS$) and the
partition function is written as a sum of basin partition functions.
In the thermodynamic limit, the system free energy $F(T)$ is written
as\cite{stillinger,costantV}

\begin{equation}
F(T) = - T  S_{conf}(e_{IS}(T)) +  f_{basin}(T, e_{IS}(T))
\label{eq:freee}
\end{equation}

\noindent
where $- T S_{conf}(e_{IS}(T)) $ account for the entropic contribution
arising from the number of basins of depth $e_{IS}$ and $f_{basin}(T,
e_{IS})$ describes the free energy of the system constrained
in one characteristic $e_{IS}$ basin.  The separation of the free
energy in two parts 
reflects the separation of time scales which
is observed in the supercooled liquid state.  Indeed, the intra-basin
dynamics is much faster ($ps$-time scale in real liquids) compared to
the inter-basin dynamics (whose time scale diverges upon cooling).  We
stress that the $IS$ expression for the liquid free energy,
Eq.~(\ref{eq:freee}) is analogous to the free-energy, derived using
the Thouless-Anderson-Palmer (TAP)\cite{tap} approximation in the
$p-$spin models, once a basin is identified with one TAP solution and
$f_{basin}(T, e_{IS})$ with the TAP free-energy.

\begin{figure}
\centerline{\psfig{figure=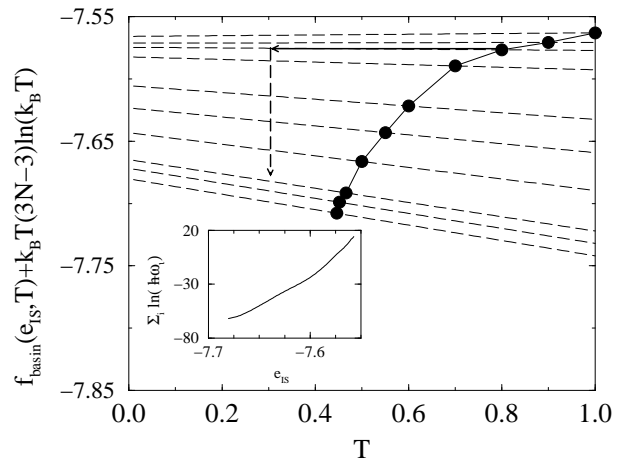,width=8.0cm,angle=0}}
\caption{
Basin free energy for the BMLJ system in harmonic approximation.
Each of the dashed line shows
$f_{basin}(e_{IS},T)$ for one specific $e_{IS}$ value
(the corresponding $e_{IS}$ value coincides with
the $T=0$ value of $f_{basin}$). In all curves, 
the $e_{IS}$-independent contribution $(3N-3)k_BTln(k_BT)$ (see 
Eq.\protect\ref{eq:hfvib})
has not been included  for clarity reasons. Filled 
circles are the equilibrium values $f_{basin}(e_{IS}(T_{eq}),T_{eq})$ . 
The arrows indicate the path followed by the system after a 
quench from $T_i=0.8$ to $T_i=0.25$
The inset shows  the 
$e_{IS}$-dependence of  $\sum_{i=1}^{3N-3} ln(\hbar \omega_i)$. The value 
of $\hbar$ is such that the dimension of $\hbar \omega$ (as well as the dimension of $k_BT$) 
are in units of the LJ potential depth\protect\cite{system}.
}
\label{fig:tap}
\end{figure}

In the case of glass-forming liquids, the intrabasin free-energy
$f_{basin}$ is usually written as $e_{IS}+ f_{vib}(T, e_{IS})$ to
separate the value of the potential energy in the minimum ($e_{IS}$)
from the vibrational free-energy contribution.  At low $T$,
$f_{vib}(T, e_{IS})$ can be calculated in the harmonic approximation,
expanding the potential energy around the $IS$ configuration, as

\begin{equation}
f_{vib}(T, e_{IS}) = k_B T \sum_{i=1}^{3N-3} 
ln(\hbar \omega_i(e_{IS})/k_B T )
\label{eq:hfvib}
\end{equation}

\noindent
where $\omega_i$ is the frequency of the $i$-th normal mode\cite{omexp}. 
If all basins had the same curvature, then $f_{vib}$ would be only
function of $T$. In the BMLJ system, basins with different depth have
different curvatures and hence $f_{vib}$ is a function of both $T$ and
$e_{IS}$.  Fig.~\ref{fig:tap} shows the $T$-dependence of
$f_{basin}(T,e_{IS})$ in the 
harmonic approximation, for different $e_{IS}$
values.  Curves for different $e_{IS}$ values are not parallel, since
the basin curvatures depend on the basin depth, as shown in the inset
of the same figure.

In equilibrium, at each temperature $T_{eq}$, the system populates
basins of different depth $e_{IS}(T_{eq})$\cite{nature}. The
$e_{IS}(T_{eq})$ value is fixed by the condition of $F$ being a
minimum, i.e. by
\begin{equation}
{{\partial F} \over { \partial e_{IS} }} = 
-T_{eq} {{ \partial S_{conf}(e_{IS}) }\over {\partial e_{IS} } }  +
{{ \partial f_{basin}(T_{eq},e_{IS}) }\over { \partial e_{IS} }} 
=0
\label{eq:teq}
\end{equation}
where the term ($ {{\partial f_{basin} }\over { \partial e_{IS} }}$)
describes the free energy change associated to the change in the basin shape
with $e_{IS}$ and the term (${{ \partial S_{conf} }\over
{\partial e_{IS} } }$) describes the entropic change associated to the
different basin degeneracy. As shown in Ref.\cite{nature}, the onset
of slow dynamics correlates with a sharp drop in the $T$-dependence of
$e_{IS}$.  This observation holds for all model systems studied so
far.  For the BMLJ system, the $e_{IS}$ basins which are populated at
the different temperatures (in the $T$-range accessible to equilibrium
simulations) are indicated in Fig.~\ref{fig:tap}. As indicated by
Eq.~(\ref{eq:teq}), the $T$-dependence of $e_{IS}$ arises from a 
balance between the change of basin free energy and the change in
configurational entropy.  Of course, the monotonic relation
$e_{IS}(T_{eq})$ can be inverted to give the temperature at which the
equilibrium system populates basins of depth $e_{IS}$,
$T_{eq}(e_{IS})$\cite{epl}.

The analysis of the non-equilibrium dynamics presented in this Letter
is based on the assumption that the separation of intrabasin
and interbasin time scales--- which characterize equilibrium
supercooled liquid states --- retains its validity during aging. In
particular, we assume that after a temperature jump from $T_i$ to
$T_f$, the vibrational intrabasin dynamics thermalizes quickly to the
thermostat value $T_f$. The thermalization of the entire system is
instead very slow, requiring a search for the low energy minima.  Such
a search in configuration space, which is the essence of the aging
phenomenon in liquids, is so slow that quasi-equilibration on basins
of depth $e_{IS}$ might be faster than the decrease of the $e_{IS}$
value.

\begin{figure}
\centerline{\psfig{figure=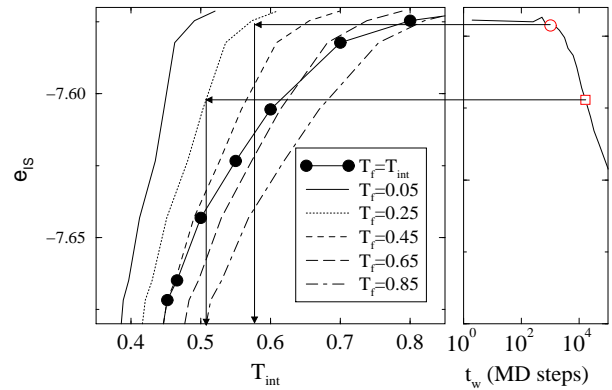,width=8.0cm,angle=0}}
\caption{Left: Solutions of  Eq.~(\ref{eq:teff}) for several  $T_f$ 
values         for the studied BMLJ system. 
Right: $e_{IS}$ as a function of time, following a $T$-jump from
$T_i=0.8$ to $T_f=0.25$. The arrows show graphically the procedure which connects the $e_{IS}(t)$ value to the $T_{int}$ value, once $T_{f}$ is known.  
Note that if the curvature of the basins were independent on $e_{IS}$, then
curves for different $T_f$ would all coincide with $T_{eq}(e_{IS})$ 
(filled symbols).
}
\label{fig:teff}
\end{figure}

In the case of BMLJ, the out-of-equilibrium dynamics following a
$T$-jump have been recently studied\cite{kobbarrat,epl}.
Ref.\cite{epl} indeed suggested the possibility that the equilibration
in configuration space proceeds via quasi-equilibrium steps.  The
proposed equilibration process is schematically depitched with arrows
in Fig.~\ref{fig:tap}, for the case of a jump from $T_i=0.8$ to
$T_f=0.25$.  In a time much shorter than the any basin change,
$f_{basin}$ assumes the value characteristic of the final temperature
(full line arrow in Fig.~\ref{fig:tap}). The fast equilibration of the
intra-basin degrees of freedom is then followed by a much slower
process (dashed arrow in Fig.~\ref{fig:tap}) during which the system
populates deeper and deeper $e_{IS}$ levels.  For the BMLJ case,
$e_{IS}(t)$ after a $T$-jump
is reported in Fig.~\ref{fig:teff}.

If the hypothesis of quasi-equilibrium is correct, then we can ask
which is the value of the internal temperature $T_{int}(e_{IS},T_f)$
selected by the system when it is populating basins of 
depth $e_{IS}$\cite{temp}.
To calculate $T_{int}(e_{IS},T_{f})$, we again search for solutions of
Eq.~(\ref{eq:teq}) but,  in contrast to the equilibrium case, we
consider the value $e_{IS}$ to be fixed and solve for the unknown
temperature, obtaining


\begin{equation}
T_{int}(e_{IS},T_{f}) = { {1+ {{   \partial } \over { \partial e_{IS} }}           f_{vib} (T_f,e_{IS})     }\over{ 
{ {\partial } \over {\partial e_{IS}} }  S_{conf}(e_{IS})     }}
\label{eq:teff}
\end{equation}

\noindent
Note that, differently from Eq.~(\ref{eq:teq}), $f_{vib}$ is now
evaluated at the thermostat temperature $T_f$, since the fast
intrabasin degrees of freedom are already thermalized to $T_f$.
This expression for $T_{int}$ coincides with the expression proposed
by Franz and Virasoro\cite{silvio} in the context of $p-$spin systems,
once the basin free energy is identified with the TAP free energy.  We
note that $ {{\partial S_{conf}} \over {\partial e_{IS} } }$ can be
evaluated from equilibrium conditions Eq.~(\ref{eq:teq}), and thus,
once a model for $f_{vib}$ is chosen, $T_{int}(e_{IS},T_{f})$ can be
calculated. If $T_f$ is small (as it usually is), the harmonic
approximation for $f_{vib}$ can be confidently used. In this case,
from Eq.~(\ref{eq:teff}) and Eq.~(\ref{eq:teq}) we obtain

\begin{equation}
T_{int}(e_{IS},T_{f}) = 
 {{1+k_B T_{f}
  \sum_i {{ \partial ln [ \hbar \omega_i / k_B T_f] } \over { \partial
 e_{_{IS}}  }}
}\over{ 1+k_B T_{eq}(e_{IS}) 
  \sum_i {{ \partial ln [\hbar \omega_i/k_B T_e ] } \over { \partial
 e_{IS}  }}
}}   T_{eq}(e_{IS})
\label{eq:teff2}
\end{equation}

\noindent
Fig.~\ref{fig:teff}-left shows $T_{int}(e_{IS},T_{f})$ for the BMLJ
system.  We note that, if basin curvatures were independent on the
$e_{IS}$ value, then the derivatives in Eq.~(\ref{eq:teff2}) would be
zero. Hence, $T_{int}(e_{IS},T_{f})$ would not depend of $T_{f}$ and
$T_{int}$ would coincide with $T_{eq}(e_{IS})$. In this limit,
$T_{int}$ plays the same role as the fictive temperature introduced in
the analysis of experimental data in aging systems\cite{hodge}.
We note in passing that an important by-product of the present
approach is a free-energy expression for an out-of-equilibrium liquid,
that depends only on $e_{IS}$ and $T_{f}$. Such expression offers a
detailed interpretation of the free energy expression for glassy
systems proposed in recent years\cite{teo}.


To test the predictions of Eq.~(\ref{eq:teff}-\ref{eq:teff2}) ---
i.e. the assumption of quasi-equilibrium in the aging system --- we
study the response of the liquid, described by an Hamiltonian $H_0$,
to an external perturbation switched on at $t=t_w$.  If the
perturbation adds the term $H_P=-V_o B({\bf r^N}) \theta(t-t_w) $ to
the Hamiltonian (where $\theta(t)$ is the Heaviside step function),
linear response theory predicts that the time evolution of any
variable $A({\bf r^N})$ conjugated to $B$ is given by\cite{hansen}

\begin{equation}
\langle A(\tau) \rangle = - {{V_o}\over{k_B T}} [ 
\langle A(\tau)B(0) \rangle_0 - \langle A(0)B(0) \rangle]_0]
\end{equation}
\noindent
where $\tau \equiv t-t_w$, $\langle \cdot \cdot \cdot \rangle$ is the
ensemble average over the perturbed system ($H_0+H_P$) and $\langle
\cdot \cdot \cdot \rangle_0$ is the ensemble average over the
unperturbed system ($H_0$).  We chose $ B \equiv (\rho^{\alpha}_{\bf
k}+\rho^{\alpha*}_{\bf k})$, where $\rho^{\alpha}_{\bf k} \equiv
\sum_i^{N_{\alpha}} e^{i {\bf k \cdot r}_i^{\alpha} }/\sqrt(N) $ is
the Fourier transform component of the density of $\alpha$ particles
at wavevector ${\bf k}$, and study the response of $A \equiv
\rho^{\alpha}_{\bf k}$.  In this case, $\langle A(t) B(0) \rangle_0$
coincides with the dynamical structure factor $S_{\bf
k}^{\alpha\alpha}(t) \equiv \langle \rho^{\alpha}_{\bf k}(t)
\rho^{\alpha*}_{\bf k}(0) \rangle_0 $. Thus, with the present choice of
$A$ and $B$, linear response theory predicts \cite{hansen}
\begin{equation}
\langle  \rho^{\alpha}_{\bf k}(\tau)\rangle  = - {{V_o}\over{k_B T}} [ S_{\bf k}^{\alpha\alpha}(\tau)-S_{\bf k}^{\alpha\alpha}(0)]
\label{eq:fd}
\end{equation}
\noindent
Eq.~(\ref{eq:fd}) --- also referred as fluctuation-dissipation
theorem --- is particularly relevant for our purposes, since it
predicts that the response of the system is proportional to $T^{-1}$
and thus offers an {\it independent} way to confirm the validity of
the quasi-equilibrium hypothesis. Indeed, the quasi-equilibrium
hypothesis predicts that for short times (i.e.  in the time region
where the correlation function assumes values between the value in
zero and the plateau value) the relation between correlation and
response, Eq.~(\ref{eq:fd}), should be controlled by $V_o/k_BT_f$, since
the intrabasin dynamics is probed. Similarly, for long times (i.e.  in
the time region where the correlation function assumes values smaller
than the plateau value) the relation should be controlled by
$V_o/k_BT_{int}$, since the inter-basin dynamics is now probed.  Thus,
we predict that switching on the perturbing field when the system is
populating basins of depth $e_{IS}$, the calculated $T_{int}$ should
coincide with the temperature at which the system responds to the
external perturbation for times longer than the vibrational dynamics.

\begin{figure}
\centerline{\psfig{figure=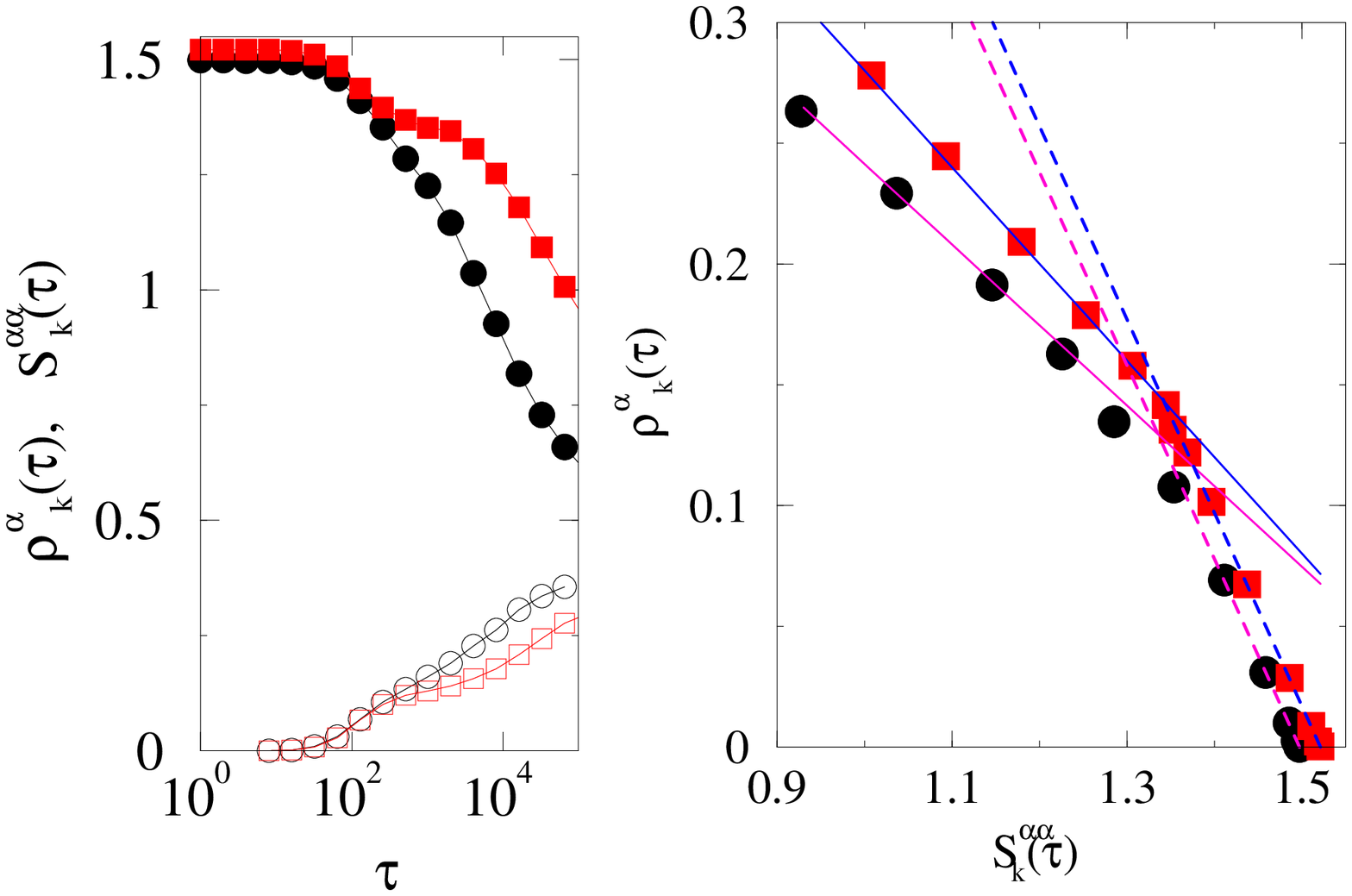,width=8.0cm,angle=0}}
\caption{Left: time dependence of the 
responce ($\rho^{\alpha}_{k}$, open symbols) 
and correlation function ($S^{\alpha\alpha}_{k}$, filled symbols)
for $t_w=1024$ (circles) and $t_w=16384$ (squares). 
Right: parametric plot (in $\tau$) of 
$\rho^{\alpha}_{k}$ vs  $S^{\alpha\alpha}_{k}$ for the two studied $t_w$.
Dashed lines have slope $V_o/k_BT_f$, thick lines have slope
$Vo/k_BT_{int}$. The two $T_{int}$ values can be read from Fig.\protect\ref{fig:teff}. The modulus of $k$ is $6.7$\protect\cite{infosim}.
}
\label{fig:cr}
\end{figure}

Fig.~\ref{fig:cr}-left shows $ S_{\bf k}^{\alpha\alpha}(\tau)$ and
$\rho_{\bf k}^{\alpha}(\tau)$ for two different $t_w$, for the BMLJ
case, with $T_i=0.8$ and $T_f=0.25$.
The reported data are averaged over 300 different
quench-realizations and over more than 60 different 
independent perturbations $H_{P}$
(but with the same $k$ modulus) for each
quench-realization\cite{infosim}.  Both $ S_{\bf
k}^{\alpha\alpha}(\tau)$ and $\rho_{\bf k}^{\alpha}(\tau)$ show the
two-step relaxation characteristic of the supercooled state, which has
been associated to the separation of intra-basin and inter-basin
motion. Fig.~\ref{fig:cr}-right shows the corresponding response
vs. correlation plots, Eq.~(\ref{eq:fd}).  At short time (intra-basin
motion) $\rho_{\bf k}(t)$ vs. $ S_{\bf k}(t)$ is linear with the
expected $T_f^{-1}$ slope, properly describing the equilibrium
condition of the vibrational dynamics with the external reservoir.  At
larger times, the intra basin motion sets-in and indeed the slope of
$\rho_{\bf k}^{\alpha}(\tau)$ vs. $ S_{\bf k}^{\alpha\alpha}(\tau)$
becomes controlled by $T_{int}^{-1}$.  As shown in Fig.~\ref{fig:cr},
the slope of the response vs. correlation plots are extremely well
predicted by Eq.~(\ref{eq:teff2}).

The good agreement between the internal temperature selected by the
system, as measured by the amplitude of the response of the aging
system to the external perturbation, and the temperature predicted
theoretically using the $IS$ formalism supports the main hypothesis on
which our analysis is based, i.e.  the validity of the
quasi-equilibrium condition.  The quality of the agreement is also a
positive test on the validity of the $IS$ formalism and of the
proposed free energy expression.  This Letter suggests that an aging
liquid, notwithstanding its out-of-equilibrium condition, can still be
described as a system in quasi-equilibrium, at the expenses of
introducing an internal temperature which is a function both of the
thermostat temperature and of the ($t$-dependent) state of the system
--- expressed by its $e_{IS}$ value. 
Eq.~(\ref{eq:teff}) more precisely defines the concept of fictive
temperature, usually defined by the experimentalist as the temperature
at which the ergodicity of the system was broken\cite{hodge}.
Eq.~(\ref{eq:teff}) clarifies that $T_{int}$ is function also of $T_f$.
We stress that the detailed analysis presented in this Letter for a
structural glass is conceptually identical to the analysis performed
in recent years to describe the out of equilibrium dynamics of
disordered spin models\cite{violation}, even though we prefer to
present our results in terms of extended validity (as opposed to
violation) of the fluctuation dissipation theorem.  Tests of the
mean-field theory on finite-size disordered $p$-spin system also
support the validity of the thermodynamic approach\cite{crisantinew}.
As noted for the case of disordered spin-systems\cite{silvio}, the
measurement of the internal temperature in aging
experiments\cite{grigera} may provide information, once properly
interpreted, on $\partial S_{conf}/ \partial e_{IS}$ i.e. on
structural properties of the system which control the equilibrium
dynamics as well.  Finally, we call attention on the fact that the
time window accessed by the numerical experiments is very different
from the experimental one. Measurements of the internal
temperature in aging experiments are very important to assess the
range of validity of the presented approach.

FS gratefully acknowledges very stimulating discussions with S. Franz. We also
thank A. Crisanti, S. Ciuchi, W. Kob, G. Parisi, G. Ruocco and F. Starr.

\begin{references}
\bibitem{debenedetti} P.~G. Debenedetti, { Metastable Liquids}
(Princeton Univ. Press, Princeton, 1997).

\bibitem{review} For recent reviews see articles in {\em Science} {\bf 267}, 
(1995) and M.D. Ediger, C.A. Angell and
S.R. Nagel, { J. Phys. Chem} {\bf 100}, 13200 (1996).

\bibitem{goetze} W. G\"otze, { J. Phys.: Condens. Matter} {\bf 11}, A1 (1999).


\bibitem{parisi} M. M\'ezard and G. Parisi, 
{ J. Phys. Cond.  Matter} {\bf 11}, A157 (1999)
M. Cardenas, S. Franz and G. Parisi, {  J. Chem. Phys.} {\bf 110}, 1726 (1999).


\bibitem{franz1} S. Franz and G. Parisi, Phys. Rev. Lett. {\bf 79}, 2486 (1997).
\bibitem{coniglio} A.Coniglio, A. de Candia, A. Fierro and M. Nicodemi
{ J. Phys.: Condens. Matter} {\bf 11}, A167 (1999). 

\bibitem{teo} Th. M. Nieuwenhuizen,
{ Phys. Rev. Lett.} {\bf 80}, 5580 (1998); {\it ibid.} {\bf 79},
1317 (1997).

\bibitem{speedy} R. Speedy,
 { J. Phys.: Conds. Matter } {\bf 10}, 4185 (1998);  R. Speedy,
J. Phys. Chem. B {\bf 103}, 4060 (1999).

\bibitem{heuer} A. Heuer, { Phys. Rev. Lett.} {\bf 78}, 4051 (1997);
S. Buechner and A. Heuer,
{ Phys. Rev. E.} {\bf 60}, 6507 (1999).

\bibitem{thomas}
T. Schr\"oder, S. Sastry, J. Dyre and S. Glotzer, J. Chem. Phys. {112},9834 (2000);

\bibitem{barbara} B. Coluzzi, Ph.D Thesis, University of Roma La Sapienza (1999). B. Coluzzi, P. Verrocchio and G. Parisi,  Phys. Rev. Lett.
{\bf 84}, 306 (2000)


\bibitem{prlentro}  F. Sciortino, W. Kob and P. Tartaglia,
Phys. Rev. Lett.  {\bf 83}, 3214 (1999).

\bibitem{srinew} S. Sastry, Phys. Rev. Letts. (in press) (2000).






\bibitem{goldstein} M. Goldstein, { J. Chem. Phys.} {\bf 51}, 3728 (1969).

\bibitem{kob} W. Kob and H.C. Andersen, {  Phys. Rev. Lett.} {\bf 73},
1376, (1994); {  Phys. Rev. E} {\bf 51}, 4626 (1995); 
{ Phys. Rev.} E {\bf 52}, 4134 (1995).
M. Nauroth and W. Kob, { ibid.} {\bf55}, 675 (1997);
T. Gleim, W. Kob, and K. Binder, {Phys. Rev. Lett.} {\bf 81},
4404 (1998).

\bibitem{system} 
The system is composed by $N=1000$ particles of a Lennard Jones (LJ)
binary (80:20) mixture with interaction parameters chosen so that the
liquid does not crystallize and does not demix.  
The LJ parameters ($\epsilon,\sigma$) of the majority component are 
the units of energy and length used in this Letter.
This model has been
extensively studied in the past\protect\cite{kob} since it does not crystallize. It has been shown that
the atomic dynamics is well described by mode coupling 
theory\protect\cite{goetze}, with a
critical temperature $T_c$ equal to $0.435$\protect\cite{kob}.

%

\bibitem{stillinger} 
F.H. Stillinger and T.A. Weber,  { Phys. Rev. A} {\bf 25}, 978 (1982); 
{ Science} {\bf 225}, 983 (1984). 
F. H. Stillinger, { Science}, {\bf 267}, 1935 (1995).

\bibitem{costantV} The adopted $IS$ formalism is
for the NVT ensemble. Generalization to NPT does not pose particular
problems. The description we present in this Letter
refers to a constant $V$. Thus, it is based on
one internal parameter (in the language of Davies and Jones
[R.O. Davies and G.O. Jones, { Adv.  in Physics} {\bf 2}, 370 (1953).]),
which we identify with $e_{IS}$. 

\bibitem{tap} D.J. Thouless, P.W. Anderson and R.G. Palmer, 
{ Phil. Mag.}, {\bf 35} 593 (1977). 
A. Crisanti, H. Horner and H.J. Sommers, { Z. Phys.} B {\bf 92}, 257 (1993).


\bibitem{omexp} $\omega_i$ can be calculated by diagonalizing the Hessian  matrix
evaluated at the $IS$ configuration (a local minimum of the potential energy). 

\bibitem{nature}S. Sastry, P. G. Debenedetti, and F. H. Stillinger, 
{\it Nature} {\bf 393}, 554 (1998).
See also H. Jonsson and 
H.C. Andersen, {Phys. Rev. Lett.} {\bf 60}, 2295 (1988).

\bibitem{epl} W. Kob, F. Sciortino, and P. Tartaglia,
{Europhys. Lett.} {\bf 49}, 590 (2000).

\bibitem{kobbarrat} J.L. Barrat and W. Kob, 
{\em Europhys. Lett.} {\bf 49}, 590 (2000); 
W. Kob and J.L. Barrat,{ Europhys. J.} B (in press).


\bibitem{temp}  L. Cugliandolo, K. Kurchan, L. Peliti, Phys. Rev. E,
{\bf 55}, 3898 (1997).

\bibitem{silvio} A detailed interpretation of this relation is given in
S. Franz, M. A. Virasoro, { J. Phys.} A {\bf 33}, 891 (2000).

\bibitem{hodge} For a review see for example I, Hodge, { J. Non-Cryst. Solids}
{\bf 131}, 435 (1991); {\bf 169}, 211 (1994).

\bibitem{hansen} J.P. Hansen and I.R. McDonald, {\em Theory of Simple
Liquids} (Academic Press, London, 1986), 2nd Edition.

\bibitem{infosim} We have equilibrated 300 independent configurations
at $T=0.8$ in the $NVT$ ensemble (Nos\'e-Hoover thermostat). Each of
the configuration has been quenched to $T_f=0.25$ by changing at $t=0$
the thermostat temperature to $T_f$. The thermostat constant is chosen
in such a way that, within 1000 MD steps, the average kinetic energy
thermalizes to $T_f$.
Configurations after $t_w=1024$ and $t_w=16386$ MD steps have been
saved. Each of the saved configurations has been used as starting
configuration for 60 independent simulations (each of them lasting $10^5$
MD steps) with
Hamiltonian $H+H_P$ ($V_o=0.2)$, each of them with a different ${\bf k}$
vector.  $|\bf{k}| $ has been constrained to the value 6.7, the
location of the first minimum of $S_{\bf k}^{\alpha\alpha}$.
The total number of MD step requested in the reported calculations is thus
the total number of configurations (300) times the
number of different perturbations (60), times the number of MD step of each of them ($10^5$), times number of studied $t_w$ (2). The calculation has
requested
a cluster of 12 alpha processors running full time for about 6 months.


\bibitem{violation} L. F. Cugliandolo and J. Kurchan,
{Phys. Rev. Lett.} {\bf 71}, 173 (1993).

\bibitem{crisantinew} A. Crisanti and F. Ritort, cond-mat/9911226.
see also A. Crisanti and F. Ritort, Europhysics Lett. (in press)

\bibitem{grigera} T. S. Grigera and N. E. Israeloff, 
{Phys. Rev. Lett.} {\bf 83}, 5038 (1999).


\end{references}
\end{document}